\documentclass[letterpaper,american,preprint]{article}
\usepackage[T1]{fontenc}
\usepackage[utf8]{inputenc}
\usepackage{color}
\usepackage{babel}
\usepackage{refstyle}
\usepackage{float}
\usepackage{booktabs}
\usepackage{units}
\usepackage{amsmath}
\usepackage{amssymb}
\usepackage{graphicx}
\usepackage[authoryear]{natbib}
\usepackage{microtype}
\usepackage[unicode=true,pdfusetitle,
 bookmarks=true,bookmarksnumbered=false,bookmarksopen=false,
 breaklinks=false,pdfborder={0 0 1},backref=false,colorlinks=true]
 {hyperref}
\hypersetup{
 linkcolor=blue,citecolor=blue,urlcolor=blue,filecolor=blue}

\makeatletter

\AtBeginDocument{\providecommand\figref[1]{\ref{fig:#1}}}
\AtBeginDocument{\providecommand\eqref[1]{\ref{eq:#1}}}
\AtBeginDocument{\providecommand\subsecref[1]{\ref{subsec:#1}}}
\AtBeginDocument{\providecommand\Figref[1]{\ref{Fig:#1}}}
\AtBeginDocument{\providecommand\Eqref[1]{\ref{Eq:#1}}}
\AtBeginDocument{\providecommand\tabref[1]{\ref{tab:#1}}}

\providecommand{\tabularnewline}{\\}
\floatstyle{ruled}
\newfloat{algorithm}{tbp}{loa}
\providecommand{\algorithmname}{Algorithm}
\floatname{algorithm}{\protect\algorithmname}
\RS@ifundefined{subsecref}
  {\newref{subsec}{name = \RSsectxt}}
  {}
\RS@ifundefined{thmref}
  {\def\RSthmtxt{theorem~}\newref{thm}{name = \RSthmtxt}}
  {}
\RS@ifundefined{lemref}
  {\def\RSlemtxt{lemma~}\newref{lem}{name = \RSlemtxt}}
  {}

\usepackage{babel}
\AtBeginDocument{\providecommand\Figref[1]{\ref{Fig:#1}}}\AtBeginDocument{\providecommand\Eqref[1]{\ref{Eq:#1}}}

\usepackage{neurips_2020}

\usepackage{url}
\usepackage{amsfonts}
\usepackage{nicefrac}

\usepackage{placeins}
\usepackage{algorithm}
\usepackage{algpseudocode}

\author{%
Sandra Nestler$^{1,2,5}$, Christian Keup$^{1,5}$,
David Dahmen$^{1}$,\\ {\bf Matthieu Gilson$^{1,4}$, Holger Rauhut$^{2}$, Moritz Helias$^{1,3}$}\\
$^{1}$Institute of Neuroscience and Medicine (INM-6), J\"ulich Research Centre, J\"ulich, Germany\\
$^{2}$Mathematics of Information Processing, RWTH Aachen University, Aachen, Germany\\
$^{3}$Department of Physics, RWTH Aachen University, Aachen, Germany \\
$^{4}$Universitat Pompeu Fabra, Barcelona, Spain\\
$^{5}$RWTH Aachen University, Aachen, Germany \\
\texttt{\{s.nestler,c.keup,d.dahmen,m.gilson,m.helias\}@fz-juelich.de,}\\ \texttt{rauhut@mathc.rwth-aachen.de} \\
}

\addto\captionsenglish{\renewcommand{\algorithmname}{Algorithm}}

\makeatother

\begin{document}
\global\long\def\TS{T_{S}}%
\global\long\def\Tn{T_{n}}%
\global\long\def\TM{T_{M}}%
\global\long\def\Siginv{\Sigma^{-1}}%
\global\long\def\TRO{T}%
\global\long\def\d{\mathrm{d}}%
\global\long\def\diag{\mathrm{diag}}%
\global\long\def\defeq{\vcentcolon=}%
\global\long\def\tr{\mathrm{tr}}%
\global\long\def\lcum{\langle\!\langle}%
\global\long\def\rcum{\rangle\!\rangle}%
\global\long\def\T{\mathrm{T}}%
\global\long\def\softM{\kappa_{\eta}}%

\title{Unfolding recurrence by Green's functions for optimized reservoir
computing}
\maketitle
\begin{abstract}
Cortical networks are strongly recurrent, and neurons have intrinsic
temporal dynamics. This sets them apart from deep feed-forward networks.
Despite the tremendous progress in the application of feed-forward
networks and their theoretical understanding, it remains unclear how
the interplay of recurrence and non-linearities in recurrent cortical
networks contributes to their function. The purpose of this work is
to present a solvable recurrent network model that links to feed forward
networks. By perturbative methods we transform the time-continuous,
recurrent dynamics into an effective feed-forward structure of linear
and non-linear temporal kernels. The resulting analytical expressions
allow us to build optimal time-series classifiers from random reservoir
networks. Firstly, this allows us to optimize not only the readout
vectors, but also the input projection, demonstrating a strong potential
performance gain. Secondly, the analysis exposes how the second order
stimulus statistics is a crucial element that interacts with the non-linearity
of the dynamics and boosts performance. 
\end{abstract}

\section{Introduction}

Trained neural networks today form an integral component of data science.
Widely used approaches comprise deep neural networks \citep{lecun2015yoshua}
that typically employ time-independent mappings by hierarchical structures
with mostly feed-forward connections. In contrast, recurrent neural
networks, which follow more closely their biological counterparts
in the brain, have units with intrinsic temporal dynamics that allow
natural processing of time-dependent stimuli. The interplay of recurrence
and non-linearity in such networks renders their analysis challenging.
There is large interest in understanding the basis for their computational
abilities. Reservoir computing, as originally introduced via Echo
State Networks \citep{Jaeger01_echo} and Liquid State Machines \citep{maass2002real},
is one approach that takes recurrence of connections and temporal
dynamics into account. Signals are here mapped into a high dimensional
space spanned by a large number of typically randomly connected neurons,
on which a linear readout is trained. The network thereby acts like
a kernel in a support vector machine \citep{vapnik1998support,cortes1995support}.
The training can be combined with a feedback of the readout signal
to effectively modify also the recurrent connections \citep{sussillo2009generating,depasquale2018full}.
The gradient of an arbitrary loss function for these models can be
computed memory efficiently via ordinary differential equations \citep{chen2018neural}.
Although recurrent models lately have become more and more complex
\citep{hochreiter1997long,cho2014learning,collins2016capacity}, they
remain highly similar to simple reservoirs in terms of the learned
neural representations \citep{maheswaranathan2019universality}. Furthermore,
it has been extensively studied how the performance of the reservoir
depends on the properties of the recurrent connectivity; the edge
of chaos has been found as a global indicator of good computational
properties \citep{bertschinger2005edge,toyoizumi2011beyond}. However,
the interplay of recurrence and non-linearities may, depending on
the statistical features of the input data, offer optimal settings
that are not described by such global parameters alone.

We here set out to systematically analyze the kernel properties of
recurrent time-continuous networks in a binary time series classification
task. We show how the high-dimensional and non-linear transformation
implemented by the network can be used to selectively extract differences
in the statistics between a pair of input classes. To this end, we
analyze the mapping between the input data distribution and the shape
and linear separation of the resulting network states, which uniquely
determine the optimal readout projection. In state-of-the-art reservoir
computing, the projection of the stimuli into the network is mostly
carried out with random weights. To the contrary, we here show that
the classification performance crucially depends on the input projection;
random projections consistently lead to significantly sub-optimal
performance, whereas an optimal input projection exploits the mode
landscape of the reservoir to obtain an advantageous configuration
of the resulting distribution of network states. We derive a method
to jointly optimize both projections in a system of linear units and
generalize these results to non-linear networks. To this end, we employ
a perturbative approach that transforms the non-linear recurrent network
into an effective feed-forward structure. The analytical expressions
expose how the network dynamics separates a priori linearly non-separable
time-series. We find that even weak non-linearities can significantly
boost the separability of network states if the linear separability
of the stimuli is low.

\section{Setup}

\begin{figure}
\begin{centering}
\includegraphics[width=0.8\linewidth]{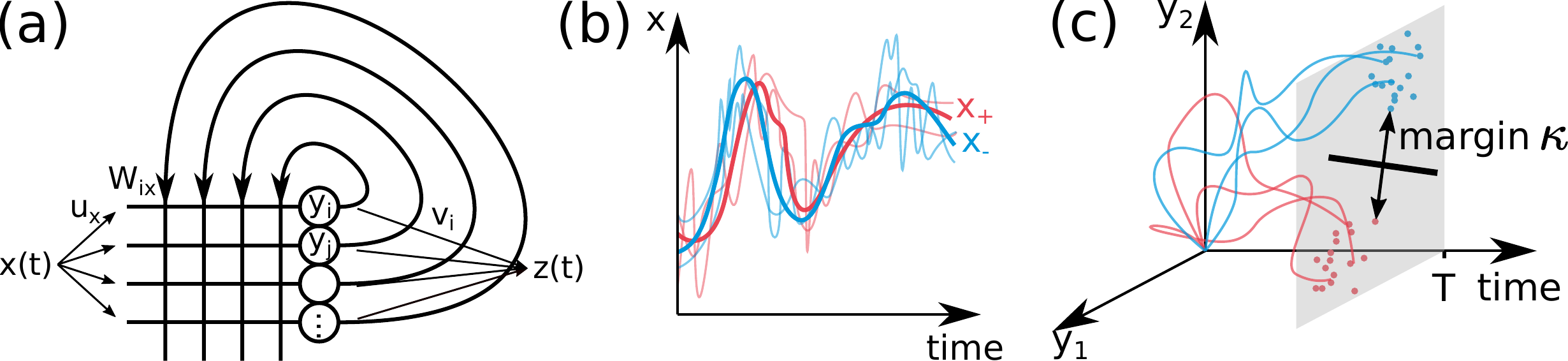} 
\par\end{centering}
\caption{\textit{Binary classification with recurrent dynamics.} (a) A neural
network with random connectivity $W$ is stimulated with an input
$x(t)$ via an input vector $u$ (left). A linear readout with weights
$v$ transforms the high dimensional state into a scalar quantity
$z(t)$. (b) Time course of sample stimuli (colored thin curves) from
two different classes (red, blue; thick curves: class average). In
this example, classes differ mainly in fluctuations. (c) Responses
of the network follow high-dimensional trajectories (colored curves,
only two dimensions $y_{1},y_{2}$ shown for conceptual clarity).
At readout time $T$, the samples form clouds of states, indicated
by points in the readout time plane. Classification places a decision
plane between the classes. The margin $\kappa$ is the smallest distance
between the states and the plane.\label{fig:Reservoir-Computing-Scheme}}
\end{figure}

We consider a reservoir model shown in \figref{Reservoir-Computing-Scheme}(a):
A time-dependent input function $x(t)$ is projected into the $N$-dimensional
neuron space with input projection $u\in\mathbb{R}^{N}$. This signal
reverberates in the network through continuous interactions via recurrent
connections $W\in\mathbb{R}^{N\times N}$ as well as sustained external
stimulation, leading to a neural trajectory $y(t)\in\mathbb{R}^{N}$
that is described by the first-order differential equation \citep{Sompolinsky88_259}
\begin{equation}
(\tau\partial_{t}+1)\,y_{i}(t)=\sum_{j}W_{ij}\phi(y_{j}(t))+u_{i}x(t),\label{eq:eom_network}
\end{equation}
where $\phi$ is the (non-linear) gain function of the neurons. The
network activity is read out linearly by the one-dimensional projection
$z(t)=v^{\T}y(t)$, obtained with readout vector $v\in\mathbb{R}^{N}$.
We here consider fixed realizations of i.i.d. weights $W_{ij}\sim\mathcal{N}(0,\frac{g^{2}}{N})$
denoting the connections from neurons $j$ to neurons $i$ and aim
towards a joint optimization of input and readout projections $u$
and $v$, respectively. In general, the existence of optimal projection
vectors allows one to first define and second study the performance
of the recurrent reservoir itself. Thus, common methods for optimizing
recurrent connectivity can be combined with our algorithm to study
and improve the kernel properties of a reservoir network, eliminating
variability of performance caused by sub-optimal input and readout
projections.

Consider inputs from two classes $+$, $-$ defined by their underlying
statistics, for example their mean trajectories and fluctuations,
as shown in \figref{Reservoir-Computing-Scheme}(b). The network transforms
the differences across classes into distinct sets of network states
$y_{\pm}(t)$, which form extended clouds in state space due to intra-class
variability (\figref{Reservoir-Computing-Scheme}(c)). For classification,
the network space is divided by a hyperplane into one region for each
class. Position and orientation of this plane are modified by the
training algorithm of the readout projection $v$, the hyperplane's
normal vector. The margin, the distance between the plane and the
sample state closest to it, is hereby a typical optimization objective
\citep{vapnik1998support,cortes1995support}.

\section{Linear Networks\label{subsec:LINEAR-NETWORKS}}

To introduce the concepts, we first investigate the benefits of optimized
input projections for linear reservoirs, where $\phi(y)=y$ in \eqref{eom_network}.
The linear equation of motion has the Green's function \citep{Risken96}
\begin{equation}
G^{(1)}(t,t^{\prime})=H(t-t^{\prime})\,\frac{1}{\tau}\exp\Big[-(\mathbb{I}-W)\frac{t-t^{\prime}}{\tau}\Big],\label{eq:Greensfunction}
\end{equation}
where $H$ is the Heaviside function. The state of neuron $i$ at
time point $t$ is then given by 
\begin{equation}
y_{i}(t)=\sum_{p}\int_{-\infty}^{\infty}\d t^{\prime}\,G_{ip}^{(1)}(t,t^{\prime})\,u_{p}\,x(t^{\prime}).\label{eq:solution_linear_Greens}
\end{equation}
The margin between classes of stimuli with class labels $\zeta_{\nu}\in{\pm1}$
\begin{equation}
\kappa(u,v)=\min_{\nu}(\zeta_{\nu}v^{\T}y^{u,\nu}),\label{eq:def_margin}
\end{equation}
where $v$ has unit length, constitutes a measure to be optimized
to increase generalization performance. We here denote by $y^{u,\nu}$
the network response to stimulus $x^{\nu}$ projected via input vector
$u$, and we assumed that the separating hyperplane passes through
the origin. This choice is adequate for the stimulus set employed
below. Shifting the plane off the origin can be accounted for by incorporation
of a threshold. The margin $\kappa$ depends on both the input projection
$u$ and the readout $v$. For a given set of training data, its maximum
is uniquely defined by the support vector machine algorithm \citep{vapnik1998support,cortes1995support}.
For the joint optimization of input and readout projections we pursue
here, we use this objective as the basis to derive analytically tractable
approximations.

For generality of the optimal projection vectors and analytical insight,
it is advisable to replace the minimum function in \eqref{def_margin}
by a differentiable approximation, leading us to a soft margin which
takes into account not only the outliers, but all points weighted
by their distance to the classification plane. This has the advantage
to tolerate some outliers if this improves the distance for the majority
of samples that are closer to the plane. Here we use a soft margin
of the form \citep{lange2014applications} 
\begin{equation}
\softM(u,v)=-\frac{1}{\eta}\,\ln\Big[\sum_{\nu}\exp(-\eta\zeta_{\nu}v^{\T}y^{u,\nu})\Big].\label{eq:soft_margin}
\end{equation}
The control parameter $\eta$ regulates the importance of distances
of states close to and far from the separating hyperplane. For $\eta\rightarrow\infty$,
we recover the margin $\kappa=\lim_{\eta\rightarrow\infty}\kappa_{\eta}$.
For finite $\eta$, the soft margin becomes less sensitive to the
exact realizations of the network states than the margin $\kappa$
(\eqref{def_margin}). We show in the supplementary material (\subsecref{Convexity-of-the-soft-margin})
by Hölder's inequality that \eqref{soft_margin} is in fact concave
in $v$; it thus possesses a unique maximum in $v$. As we presume
a large number of samples representing the distribution of stimuli,
we can express the sum by an expectation value with respect to the
underlying probability distribution of $\zeta_{\nu}y^{u,\nu}$, 
\[
\softM(u,v)\rightarrow-\frac{1}{\eta}\,\ln\big\langle\exp(-\eta\zeta_{\nu}v^{\T}y^{u,\nu})\big\rangle,
\]
where we neglected an inconsequential offset. The soft margin $\softM$
has now the form of a scaled cumulant generating function \citep{Gardiner85,touchette2009large};
its Taylor expansion until the second cumulant of $\zeta_{\nu}y^{u,\nu}$
thus reads 
\begin{align}
\softM(u,v) & \approx v^{\T}M^{u}-\frac{1}{2}\eta\,v^{\T}\Sigma^{u}\,v,\label{eq:soft_margin_approx_Gaussian}
\end{align}
where $M_{i}^{u}:=\langle\zeta_{\nu}y_{i}^{u,\nu}\rangle$ is the
average separation vector between the center of the clouds and the
decision plane and $\Sigma_{ij}^{u}:=\langle(\zeta_{\nu}y_{i}^{u,\nu})\,(\zeta_{\nu}y_{j}^{u,\nu})\rangle-M_{i}^{u}M_{j}^{u}$
the covariance matrix. The two terms have counteracting effects on
the soft margin. The decomposition of the soft margin into cumulants
of labeled network states shows a suppression of cumulants of order
$k$ by a factor $\frac{1}{k!}$. It is also geometrically plausible
that lower order cumulants are more important than higher orders;
they describe the rough shape of the state clouds. Stopping after
second order amounts to a Gaussian approximation of the state clouds.
Alternatively, one can regard \eqref{soft_margin_approx_Gaussian}
as classification by linear discriminant analysis if the two sample
classes are of equal size \citep{minasny2009elements}; when further
assuming Gaussianity and equal variance of the two classes, this is
identical to Fisher linear discriminant analysis. From now on, the
term soft margin will refer to \eqref{soft_margin_approx_Gaussian}.

Since a linear gain function $\phi(y)=y$ imposes a linear relationship
between network inputs and outputs (\eqref{solution_linear_Greens}),
each cumulant of the network state depends only on the corresponding
cumulant of the stimulus. Separation between the classes is thus linearly
related to the difference between mean stimuli of the two classes.
In contrast, in non-linear networks higher order cumulants also contribute
to the separation between the classes.

Optimization of the soft margin can be done for arbitrary input signals.
However, since \eqref{soft_margin_approx_Gaussian} only depends on
the mean and covariance of network outputs, it is sufficient for linear
networks to regard stimuli as coming from a Gaussian distribution.
As an example, in the following the stimuli are furthermore taken
as step-wise constant, accounting for a finite temporal resolution
$\Delta t$ that would typically appear in a practical application.
We therefore replace the dependence on the stimulus time $t^{\prime}$
in the Green's function by the index $n$, where $t_{n}=n\,\Delta t$,
defining $G_{ipn}^{(1)}(t):=\int_{t_{n}}^{t_{n+1}}G_{ip}^{(1)}(t,t^{\prime})\,dt^{\prime}$.
Without loss of generality we can assume the distribution of stimuli
to be of the form 
\begin{align}
x^{\pm} & \propto\mathcal{N}(\pm\mu,\psi\pm\chi),\label{eq:Gaussian_distribution_input}
\end{align}
where $\mu\in\mathbb{R}^{T/\Delta t}$ and $\psi,\chi\in\mathbb{R}^{T/\Delta t\times T/\Delta t}$.
A potential offset in the mean could be absorbed by a corresponding
threshold in equations (\ref{eq:def_margin}) - (\ref{eq:soft_margin_approx_Gaussian})
and different covariances $C^{\pm}$ are included by setting $\psi:=\frac{1}{2}(C^{+}+C^{-})$
and $\chi:=\frac{1}{2}(C^{+}-C^{-})$. It is straightforward to then
compute the average separation at time $T$ 
\[
M_{i}^{u}=\sum_{p,n}\,G_{ipn}^{(1)}(T)\,u_{p}\mu_{n}
\]
and the covariance 
\[
\Sigma_{ij}^{u}=\sum_{n,m,p,q}G_{ipn}^{(1)}(T)\,G_{jqm}^{(1)}(T)\,u_{p}u_{q}\,\psi_{nm}.
\]
The soft margin (\eqref{soft_margin_approx_Gaussian}) is thus quadratic
in both the input projection $u$ and the readout vector $v$ and
therefore simple to optimize with respect to either of them. Hereby,
we require both projection vectors to be normalized. For the readout,
this ensures a meaningful calculation of the margin. For the input
projection, this fixes the amplitude of the driving signal (cf. \subsecref{NONLINEAR-NETWORK}
and supplementary material (\subsecref{Appendix})). A constrained
optimization follows with the method of Lagrange multipliers by computing
the stationary points of 
\begin{align}
\mathcal{L}(u,v):= & \softM(u,v)+\lambda_{u}(\|u\|^{2}-1)+\lambda_{v}(\|v\|^{2}-1)\label{eq:def_opt_with_Lagrange}
\end{align}
with $\lambda_{u/v}<0$ (see supplementary material, \subsecref{Constrained-optimization}).
This equation can be maximized by alternating fixed-point iteration.
In case $\mu=0$, the soft margin is a quadratic form and finding
the optimal projection vectors reduces to an eigenvalue problem. A
detailed description of the optimization process is given in the supplementary
material, \subsecref{Appendix}.

\begin{figure}
\begin{centering}
\includegraphics[width=1\textwidth]{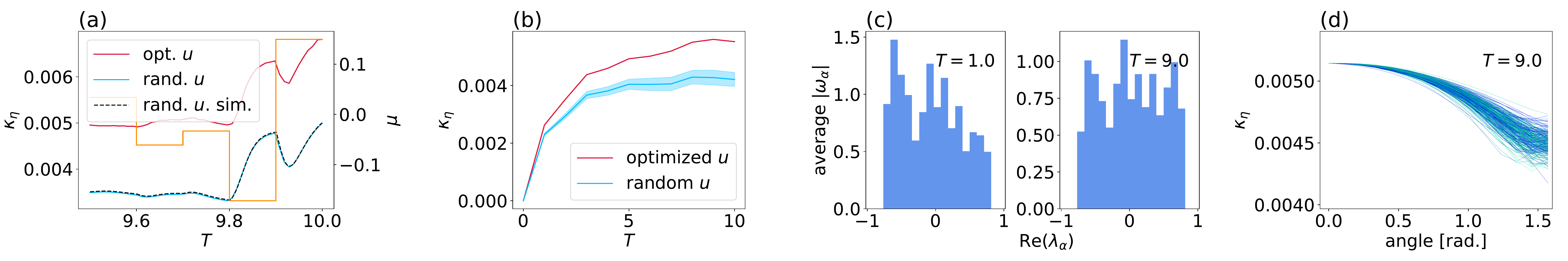} 
\par\end{centering}
\caption{\textit{Optimization of the input and readout projections in a reservoir
of linear units.} (a) Random network with $N=100$ neurons, $\tau=0.25$
and fixed connectivity $W$ ($W_{ij}\sim\mathcal{N}(0,\frac{g^{2}}{N})$,
$g=0.9$) is stimulated by step-wise constant stimuli with a mean
separation of $\mu=\frac{1}{2}(x_{+}(t)-x_{-}(t))$ (orange, right
scale). The soft margin (left scale) when stimulated with a random
input projection, but readout vector optimized according to \eqref{soft_margin_approx_Gaussian}
(blue: analytical solution (\eqref{solution_linear_Greens}), black
dashed: simulation result \citep{Gewaltig_07_11204}) with $\eta=10$
is shown alongside the combined optimization using $30$ optimization
steps of input and readout projection at readout time $T$ (red).
(b) The same network is stimulated with $250$ random stepwise constant
signals with optimized (red) and $250$ random (blue) input projections
each. The corresponding readout projection is chosen optimally in
either case. The soft margins, averaged over stimuli, are shown for
both cases. Colored area marks the standard deviations of $\kappa_{\eta}$
with respect to random input projections averaged over stimuli. (c)
At readout times $T=1$ and $T=9$, the optimal input projections
of the samples used in (b) are decomposed into eigenmodes of the reservoir.
Histograms show the average absolute weight $\omega_{\alpha}$ of
the modes corresponding to an eigenvalue $\lambda_{\alpha}$; real
part determines the time constant $\tau_{\alpha}$ of the mode. (d)
Soft margin for varying input projection $u$ that has the given angle
on the abscissa to the optimal direction $u^{\ast}$ at readout time
$T=9$; $u$ is oriented within a randomly chosen hyperplane. \label{fig:Optimization-of-input-output}}
\end{figure}

\Figref{Optimization-of-input-output} shows the increase in soft
margin by optimizing the input projection $u$ in the linear reservoir.
Inspecting the time span just prior to the readout time point $T$
exposes the high sensitivity of the soft margin to the readout time
point (\Figref{Optimization-of-input-output}a). The global optimum
may thus be reached at some intermediate time point, prior to the
end of the stimulus; it is possible that later steps of the stimuli
counteract the separation or disturb the favorable orientation of
the state clouds. On average over many sets of stimuli, however, the
soft margin increases towards late readout times (\Figref{Optimization-of-input-output}b),
indicating that the reverberating activity of the network can effectively
be used to accumulate evidence. Networks closer to instability with
longer time constants or a time-integrated readout may therefore be
beneficial for the performance, particularly in the example shown
in \subsecref{application-to-data-set}. Qualitatively, the dominance
of the more recent past of the stimulus, however, prevails. An average
over stimuli of the decomposition $\omega_{\alpha}=w_{\alpha}^{\T}u$
of the optimal input projections into eigenmodes $w_{\alpha}$ of
the connectivity, $w_{\alpha}^{\T}\,W=\lambda_{\alpha}w_{\alpha}^{\T}$,
is shown in \Figref{Optimization-of-input-output}(c). The information
projected on each mode thereby decays exponentially with time constant
$\tau_{\alpha}=\tau\,\big(1-\text{Re}(\lambda_{\alpha})\big)^{-1}$.
Pronounced contributions of modes with short time constants at both
early and late readout times emphasize the importance of the recent
past of the stimulus for classification. Perturbing the input projection
$u$ into random directions shows that the optimal direction is sharply
defined (\Figref{Optimization-of-input-output}(d)).

\section{Non-linear Networks\label{subsec:NONLINEAR-NETWORK}}

Classification by a linear system fails when stimuli become linearly
inseparable, because the mapping of the stimulus into the state space
of the network can only perform a linear transformation. The introduction
of a non-linear activation function qualitatively changes this result.
Interpreting the processing in the network as a kernel functional,
the space it belongs to is thus extended: to leading order in a perturbative
expansion, the mapping changes from a linear functional to a quadratic
functional; that is, a functional in which pairs of time points of
the input signal contribute to the network output at any given point
in time. These non-linear interactions render the system sensitive
to class-specific characteristics also in higher order cumulants.
The soft margin therefore profits from more contributions to the distance
$M$ and covariance $\Sigma$ of the state clouds. The approach therefore
elucidates which statistical features of the input data can be used
by the network, thus opening a door to link and compare reservoir
computing to feature-based approaches of classification.

We focus on the case where the neural gain function is explored only
in a confined area around a working point, where the non-linearity
remains small, so we expand the gain function as $\phi(y)\simeq y+\alpha y^{2}+\text{\ensuremath{\mathcal{O}(y^{3})}}$
with a small, positive parameter $0\leq\alpha\ll1$, and a small or
vanishing initial condition for $y$. In the context of biological
neural networks, the gain function represents the non-linearity experienced
by a single synaptic input on the background noise caused by the other
inputs. It is formally ontained by a Gram-Chalier expansion; an expansion
in the non-Gaussian cumulants of a nearly Gaussian distributed input.
\citep{dahmen2016correlated} and \citep{farkhooi2017complete} have
explored such expansions for binary networks and found that even the
linear order provides a good approximation of the recurrent dynamics,
as soon as the number of inputs per neuron is on the order of $50-100$.
For conceptual clarity, we here focus on the simpler case of a rate
network, but more elaborate methods are also conceivable. The corresponding
Green's function for the network can be derived from a perturbation
expansion of the corresponding network dynamics in orders of $\alpha$
as 
\begin{equation}
y(t)=y^{(0)}(t)+\alpha y^{(1)}(t)+\mathcal{O}(\alpha^{2}).\label{eq:rate-series}
\end{equation}
Inserting the ansatz (\eqref{rate-series}) into \eqref{eom_network}
separates the solution into different orders of $\alpha$. The zeroth
order, 
\begin{equation}
(\tau\partial_{t}+1)\,y_{i}^{(0)}(t)-\sum_{j}W_{ij}y_{j}^{(0)}(t)=u_{i}x(t)+\mathcal{O}(\alpha),\label{eq:leading-deq}
\end{equation}
recovers the linear system, solved by \eqref{solution_linear_Greens}.
Corrections to the dynamics can be found in higher orders in $\alpha$.
With use of \eqref{leading-deq}, the differential equation (\eqref{eom_network})
with terms up to first order in $\alpha$ simplifies as 
\begin{align}
(\tau\partial_{t}+1)\,y_{i}^{(1)}(t)-\sum_{j}W_{ij}y_{j}^{(1)}(t) & =\sum_{j}W_{ij}(y_{j}^{(0)}(t))^{2}.\label{eq:first_order_correction}
\end{align}
The first non-linear correction to the linear dynamics obeys the same
differential equation as the linear one, with the linear solution
entering the inhomogeneity in the place of $u_{i}x(t)$. Thus, $y^{(1)}$
follows with the Green's function $G^{(1)}$ (\eqref{Greensfunction})
and \eqref{first_order_correction} as 
\begin{align}
\alpha y_{i}^{(1)}(t) & =\alpha\sum_{i^{\prime},j}\int_{-\infty}^{\infty}\d t^{\prime}\,G_{ii^{\prime}}^{(1)}(t,t^{\prime})\,W_{i^{\prime}j}\,\big[y_{j}^{(0)}(t^{\prime})\big]^{2}\nonumber \\
 & =:\sum_{p,q}\int_{-\infty}^{\infty}\d s\,\int_{-\infty}^{\infty}\d s^{\prime}\,G_{ipq}^{(2)}(t,s,s^{\prime})\,u_{p}u_{q}\,x(s)x(s^{\prime}),\label{eq:G2-general}
\end{align}
where we defined the second order Green's function $G^{(2)}$ and
$y_{j}^{(0)}(t^{\prime})$ is the zeroth order solution of \eqref{leading-deq}
given by \eqref{solution_linear_Greens}. At this order, the reservoir
thus maps the input by a bi-linear functional kernel to the output.
Concerning the validity of the approximation, it must be noted that,
whereas the solution of the linear system remains well-defined also
in the linearly unstable regime, the perturbative solution of the
non-linear system built thereof (\eqref[s]{rate-series} - (\ref{eq:G2-general}))
in that case suffers from exponentially growing modes. Therefore,
we do not consider chaotic networks in our analysis, restricting the
variance of connectivity weights to $g<1$.

\begin{figure}
\begin{centering}
\includegraphics[width=0.8\textwidth]{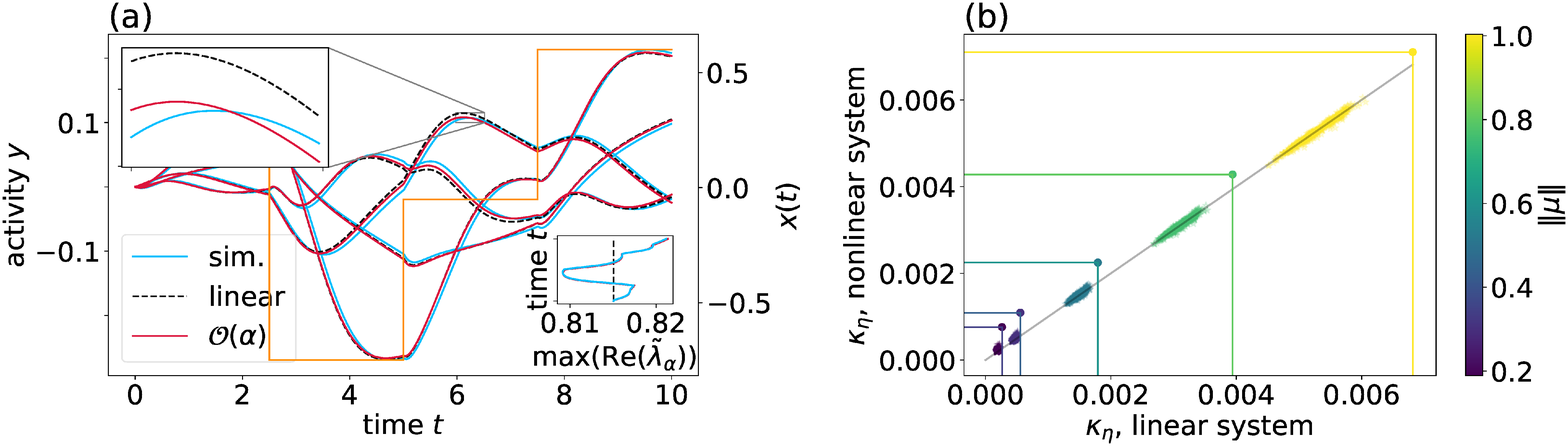} 
\par\end{centering}
\caption{\textit{Responses and soft margins in a network with small non-linearity
$\alpha$.} (a) First order approximation $\mathcal{O}(\alpha)$ of
the dynamics (red) and linear response (black dashed). Simulation
shown in blue. Left inset shows a zoom in, right inset shows time
evolution of $\max_{\alpha}(\text{Re}(\tilde{\lambda}_{\alpha}))$
for simulation (blue) and $\mathcal{O}(\alpha)$ approximation (red)
of the nonlinear reservoir together with $\max_{\alpha}(\text{Re}(\lambda_{\alpha}))$
(black dashed) of the linear reservoir. (b) Soft margins $\protect\softM$
for random (stars) and optimized input projections $u$ (dots) for
one example network realization; both cases use optimized readout
projection $v$ with respect to \eqref{def_opt_with_Lagrange}. Linear
system on x-axis, non-linear system on y-axis. Vertical and horizontal
colored lines at position of optimized solution provided as a guide.
Gray line is the angle bisector. Colors indicate $\|\mu\|$, the strength
of linear separability of the underlying stimulus distribution, where
$\|\mu\|\in\{0.19,0.30,0.52,0.76,1.0\}$ from violet to yellow. Both
$\psi$ and $\chi$ are held constant with eigenvalues of $\psi\pm\chi$
in the range $[0.3,2.2]$. Same parameters as in \figref{Optimization-of-input-output},
but with $\alpha=0.05$ for the non-linear system.\label{fig:Responses-and-soft-margin-nonlinear}}
\end{figure}

\Figref{Responses-and-soft-margin-nonlinear}(a) shows that the first
order correction in $\alpha$ approximates the dynamics of the full
system quite well. For small $\alpha$, the network is linearly stable:
the eigenvalues $\tilde{\lambda}$ of the linearized connectivity
(see supplementary material, \subsecref{Linearized-connectivity})
$\tilde{W}_{ij}=W_{ij}(1+2\alpha y_{j}(t))$ fulfill $\max(\text{Re}(\tilde{\lambda}))<1$
(\Figref{Responses-and-soft-margin-nonlinear}a, right inset). Consequently,
the difference between the linear and non-linear system is not large.
Yet, we will show that the non-linearity has a considerable impact
on the separability of inputs where the linear theory alone fails
to separate the stimuli.

Given the Green's functions $G^{(1)}$ and $G^{(2)}$, the expected
distance and covariance required for evaluation of the soft margin
in \eqref{soft_margin_approx_Gaussian} can be computed using 
\begin{equation}
\zeta_{\nu}y_{i}^{u,\nu}=\sum_{p,n}\,G_{ipn}^{(1)}(T)\,u_{p}\,\zeta_{\nu}x_{n}^{\nu}+\sum_{p,q,m,n}G_{ipqnm}^{(2)}(T)\,u_{p}u_{q}\,\zeta_{\nu}x_{n}^{\nu}x_{m}^{\nu}.\label{eq:class-generalized-states-nonlinear}
\end{equation}
The distance $M$ between the state clouds thereby receives $\mathcal{O}(\alpha)$
contributions from the first two cumulants of the stimuli, whereas
the covariance $\Sigma$ receives corrections up to $\mathcal{O}(\alpha^{2})$
from stimulus cumulants up to fourth order. Although $\mathcal{O}(\alpha^{2})$
corrections to $\Sigma$ form only a small modification to the covariance
that is otherwise determined up to $\mathcal{O}(\alpha)$, this term
is essential to guarantee its positive definiteness. A consistent
calculation of network state cumulants is therefore required for a
stable optimization algorithm. It is easy to show that all orders
in $\alpha$ of both $M$ and $\Sigma$ are affected by Gaussian distributed
stimuli. The latter is therefore the minimal example to expose cumulant-mixing
based on non-linearities. Contributions from higher order cumulants
of stimuli would not show qualitatively different effects.

\Eqref{def_opt_with_Lagrange} can thus be expressed with help of
\eqref{class-generalized-states-nonlinear}. As in the linear case
it is bi-linear in $v$, but due to $\Sigma$ it now contains terms
with third and fourth power in $u$. By the bi-linearity in $v$,
the readout projection is determined as in the linear case, only with
additional contributions to the covariance matrix and distance vector.
The optimization of the input projection, by contrast, is more challenging.
The higher powers of $u$ impede a direct solution. In our analysis,
the most reliable optimization scheme proved to be searching for a
direct solution to $\partial_{u}\mathcal{L}(u,v)=0$ given by \eqref{def_opt_with_Lagrange}
with an appropriate initial guess. More details and pseudocode can
be found in the supplementary material (\subsecref{Constrained-optimization}).

How much the choice of the input projection affects the soft margin
can be observed in \figref{Responses-and-soft-margin-nonlinear}(b).
The input and readout projections are optimized separately for a linear
and a non-linear network; they take on different optimal values for
the two reservoirs. A benefit of optimizing the input projection in
the linear reservoir only occurs for increasing strength in the mean
class difference $\mu$. For small $\mu$, the optimal direction is
dominated by the one that minimizes the effect of the noise, while
for larger $\mu$, the stimulus direction aligns such as to maximize
the mean separation of the output of the network. The situation is
clearly different in the non-linear reservoir even for the weak non-linearity
considered here: At low linear separabilities of inputs, the optimization
of the input projection in the non-linear reservoir yields a strong
relative improvement in separability of outputs, indicated by the
soft margin. For linearly well separable classes the relative improvement
with respect to the linear reservoir shrinks, while the absolute improvement
stays rather constant. Close to the information theoretic optimum
of perfectly separable classes considered in the dataset application
in \subsecref{application-to-data-set}, the benefit of non-linearities
becomes negligible. The superior performance of the weakly non-linear
system with respect to the linear system vanishes for all input separabilities
if input projections are not optimized: For random input projections,
the performance in the non-linear reservoir is on average only slightly
better, and sometimes even worse, than in the linear reservoir. The
random input projections accumulate along the identity line in \figref{Responses-and-soft-margin-nonlinear}(b),
with a center of mass slightly in the upper area.

In summary, while the weak non-linear corrections to the linear dynamics
as used here do not exploit the full computational power non-linearities
can exert, the presented routine allows us to inspect the potential
of this framework that is not apparent in classical reservoir computing
with random input projections.

\section{Application to ECG5000 dataset\label{subsec:application-to-data-set}}

We conclude the analysis with an application to a univariate temporal
classification dataset. This serves as a proof-of-concept to demonstrate
the effects of the optimization on a real-world problem and can be
regarded as a check that real data do not generally contain structural
obstacles that were not covered in the theoretical considerations.
To raise the method from the proof-of-concept level, the performance
should be systematically checked on a broader set of problems as done
for state of the art time series classifiers \citep{bagnall2017great,wang2017time},
which we leave for future work.

We here restrict the preprocessing of the data to a minimum. In this
spirit, also the free parameters $\eta$ and $\tau$ are chosen appropriately,
but not optimally. The focus lies solely on a comparison between random
and optimized input projections. This comparison is based on the classification
soft margin and accuracy in a fixed reservoir configuration. We can
then observe the effect of the optimization routine on the separation
and covariance of the state clouds.

The examined dataset is ECG5000, which is publicly available at the
UCR Time Series Classification archive \citep{chen2015ucr}, containing
5000 electrocardiograms of single heartbeat recordings. The classes
separate between five categories of healthy and diseased heartbeats.
For a binary classification, we use only samples from the two largest
classes, so that we obtained a training set consisting of 354 samples
and a testing set of 4332 samples. All stimuli were shifted and scaled
to provide classes with means $\pm\mu$ with $\|\mu\|=1$; higher
order cumulants changed accordingly. This scaling of inputs is only
performed for conceptual clarity, allowing identical network parameters
as in the previous task. Likewise, one could adapt the value of $\alpha$
according to the stimulus strength. Furthermore, for maximal performance,
a trained threshold can replace the centering of data. As a measure
of linear separability, we relate the difference of the class means
to the covariance in the direction of separation. This yields a ratio
$\nicefrac{\left\Vert \mu\right\Vert ^{2}}{\sqrt{\mu^{T}\psi\mu}}=2.6$,
which is much higher than for the artificial stimuli analyzed in \figref{Responses-and-soft-margin-nonlinear},
where the corresponding measure ranges between $0.19$ and $0.98$.

All results presented here use the same parameters as in \figref{Optimization-of-input-output}
and \figref{Responses-and-soft-margin-nonlinear}. 
\begin{table}
\caption{\textit{Quality measures for the application of the optimization scheme
to ECG5000.} Soft margin $\kappa_{\eta}$ (left) and accuracy (right)
for optimized and $50$ random input projections, averaged over $20$
different network realizations.\label{tab:ECG-results}}

\centering{}%
\begin{tabular}{lllll}
 & $\kappa_{\eta}$, linear  & $\kappa_{\eta}$, non-linear  & accuracy, linear  & accuracy, non-linear\tabularnewline
\midrule 
random $u$  & $0.182\pm0.015$  & $0.183\pm0.015$  & $(91.7\pm0.6)\,\%$  & $(91.7\pm0.7)\,\%$\tabularnewline
optimized $u$  & $0.383\pm0.026$  & $0.384\pm0.026$  & $(97.3\pm0.4)\,\%$  & $(97,3\pm0.4)\,\%$\tabularnewline
\bottomrule
\end{tabular}
\end{table}

The summary of the results in \tabref{ECG-results}, which contains
averaged results over $20$ different initializations of the recurrent
connectivity, makes evident that a maximized soft margin is accompanied
by increased accuracies. The optimized input projections outperformed
all randomly chosen ones both with respect to soft margin and accuracy.
Because of the close to perfect linear separability of the data, the
increase of soft margins and accuracies from the linear to the non-linear
reservoir is very small (see supplementary material, \subsecref{Additional-material}).
These results are as theoretically expected from \figref{Responses-and-soft-margin-nonlinear}(b)
for linearly well separable data. An application to a broader set
of real world data would be required to quantify the performance increase
in terms of accuracy also in the case of linearly less separable stimuli.

\begin{figure}
\begin{centering}
\includegraphics[width=0.8\textwidth]{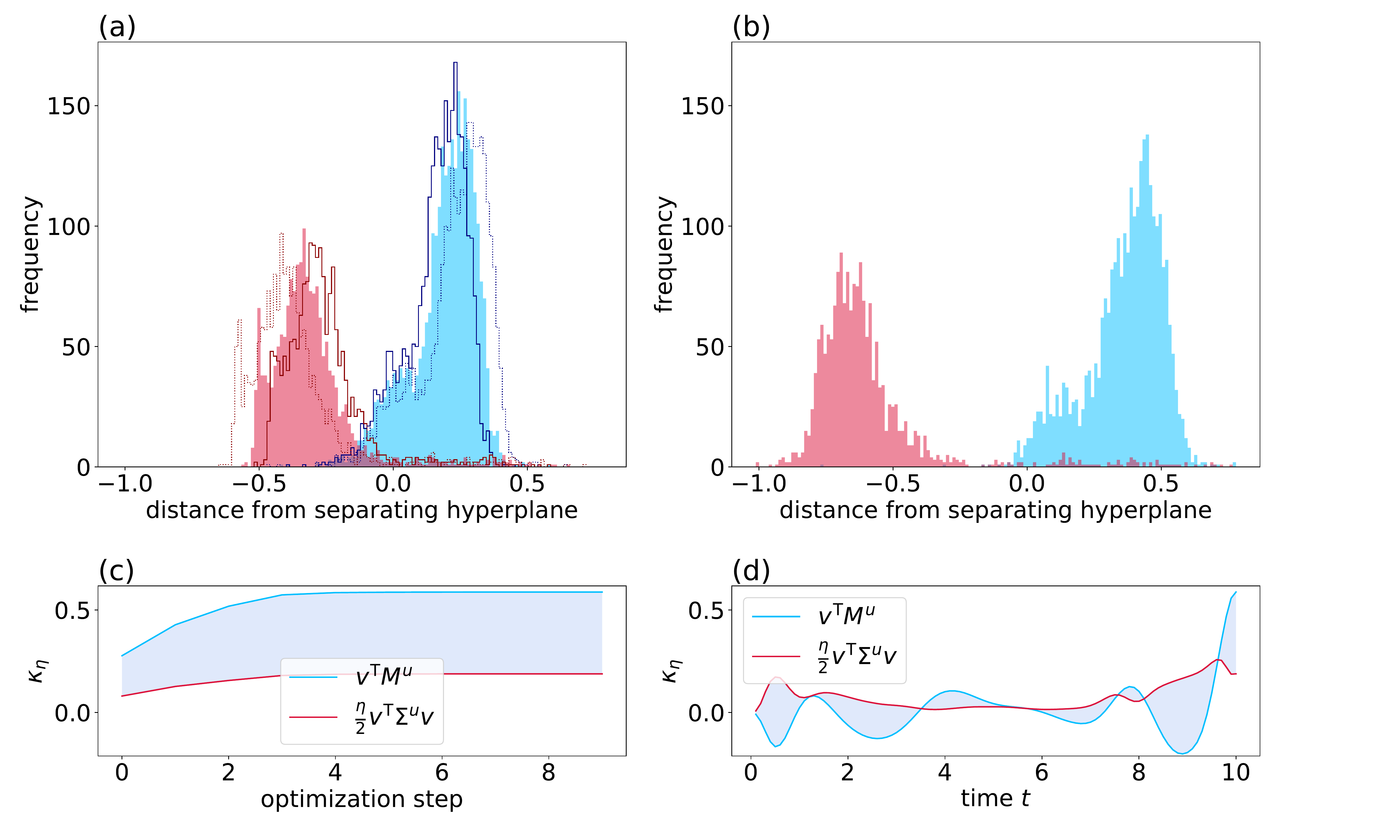} 
\par\end{centering}
\caption{\textit{Network state distribution for random and optimal input projection.}
(a) Random input projections: Colored histogram shows the network
state distribution for both classes in direction of the separating
hyperplane's normal vector (average over $50$ input projections).
Light and dark outlined histograms correspond to the input vectors
with the best and worst accuracies among the drawn samples, respectively.
All network states are based on $\mathcal{O}(\alpha)$-predictions
of the dynamics. (b) Optimized input projection: Histogram of the
network state distribution (based on the simulated responses). (c)
Evolution of distance and covariance contribution to the soft margin
$\kappa_{\eta}$ over the first 10 optimization steps at readout time.
Height of the shaded area corresponds to the resulting $\kappa_{\eta}$,
illustrating the difference between the two terms in \eqref{soft_margin_approx_Gaussian}.
(d) Evolution of distance and covariance contribution to $\kappa_{\eta}$
over simulation time for optimized input vector $u$. Height of the
shaded area corresponds to the soft margin (negative where covariance
contribution (red) exceeds distance contribution (blue) and positive
otherwise). Same network and parameters as in \figref{Responses-and-soft-margin-nonlinear}.
\label{fig:state-distribution-histogram}}
\end{figure}

A visualization of the optimization in \figref{state-distribution-histogram}
shows the increase of distance between the classes before and after
optimization (a, b), gradually increasing with the optimization step
(c). The projections being optimized for $T=10$, the two classes
become distinguishable only shortly before this readout time point
(d). The variance along the readout direction $\frac{\eta}{2}v^{\T}\Sigma^{u}v$
is hereby rather constant, while the main deviations occur due to
the separation. Increasing $\eta$ can be used to enforce smaller
dispersion of the state clouds (see supplementary material, \subsecref{Additional-material}).
The enlarged range in between the class centers with low probability
density for network states of either class facilitates a better generalization
to unknown data.

\section{Discussion}

We present an analytical approach of unrolling recurrent non-linear
networks by use of a perturbative expansion. The conceptual insight
of this step lies in a simplification of the reverberating neuronal
dynamics into an effective feed-forward structure. This approach,
which involves the first and second order Green's function of the
system, extends naturally from linear networks to non-linear ones.
The reformulation of the classification margin as a partly concave
soft margin, which has a similar form as a free energy, facilitates
the derivation of closed-form expressions to be maximized. The joint
optimization of stimulus projection and readout vector leads to a
significant increase in classification performance by tuning the network
state distribution towards a trade-off between low variability along
the direction of separation and high absolute separation. This increase
of separability is in particular observable even in only weakly non-linear
networks when the linear separability of the stimuli is low. The effect
can be fully explained by the second order Green's function that makes
the reservoir sensitive to classification features in the second order
stimulus statistics. We find that the effect of higher statistical
orders of the data are suppressed by powers in the perturbation parameter,
the non-linearity of the neuronal dynamics. But also the classification
performance of the linearly well separable dataset ECG5000 profits
significantly from the optimization. The framework presents a stepping
stone towards a systematic understanding of information processing
by recurrent random networks.

\subsubsection*{Acknowledgements}

This work was partly supported by European Union Horizon 2020 grant
945539 (Human Brain Project SGA3), the Helmholtz Association Initiative
and Networking Fund under project number SO-092 (Advanced Computing
Architectures, ACA), BMBF Grant 01IS19077A (Juelich) and the Excellence
Initiative of the German federal and state governments (G:(DE-82)EXS-SF-neuroIC002).

\subsubsection*{Broader impact}

The main motivation of this work is to provide conceptual insight.
Analytically unrolling recurrent dynamics into a (functional) Taylor
series, where coefficients are given by Green's functions, is a versatile
approach that may be used as a general purpose scheme to analyze recurrent
networks and to optimize reservoir computing. This expansion reveals
how the non-linear interactions and recurrence pick up higher order
correlations in the input statistics, quantifying how non-linear networks
provide a richer feature space than linear ones. We consider the presented
application as a proof-of-principle for optimized processing of complex
time series data. The presented application to health-related data
(heartbeat classification) hints at possible societal consequences
by providing better diagnostic tools.

\newpage{}

\appendix

\section{Supplementary material\label{subsec:Appendix}}

\subsection{Convexity of the soft-margin\label{subsec:Convexity-of-the-soft-margin}}

The soft margin $\kappa_{\eta}$ (\eqref{soft_margin}) is concave
in $v$; this follows directly from the linear appearance of $v$
in the exponent of the exponential function with Hoelder's inequality.
Hoelder's inequality states for two non-negative sequences $g_{k},h_{k}\ge0$
and for $\alpha+\beta=1$ that 
\begin{align}
\sum_{k}(g_{k})^{\alpha}(h_{k})^{\beta} & \le(\sum_{k}g_{k})^{\alpha}(\sum_{k}h_{k})^{\beta}.\label{eq:Hoelder}
\end{align}
We here follow a modified version of the argument in \citep{goldenfield1992lectures}.
It therefore follows for $\alpha+\beta=1$ that 
\begin{align*}
\kappa_{\eta}(u,\alpha v_{1}+\beta v_{2}) & =-\frac{1}{\eta}\,\ln\sum_{\nu=1}^{P}\,\exp\Big(\alpha\big[-\eta\,\zeta_{\nu}\big(v_{1}^{\T}y^{u,\nu}\big)\big]+\beta\big[-\eta\,\zeta_{\nu}\big(v_{2}^{\T}y^{u,\nu}\big)\big]\Big)\\
 & =-\frac{1}{\eta}\,\ln\sum_{\nu=1}^{P}\,\exp\Big(-\eta\,\zeta_{\nu}\big(v_{1}^{\T}y^{u,\nu}\big)\Big)^{\alpha}\,\exp\Big(-\eta\,\zeta_{\nu}\big(v_{2}^{\T}y^{u,\nu}\big)\Big)^{\beta}\\
 & \stackrel{\text{Hoelder}}{\geq}-\frac{1}{\eta}\,\ln\Bigg[\sum_{\nu=1}^{P}\exp\Big(-\eta\,\zeta_{\nu}\big(v_{1}^{\T}y^{u,\nu}\big)\Big)\,\Bigg]^{\alpha}\,\Bigg[\sum_{\nu=1}^{P}\exp\Big(-\eta\,\zeta_{\nu}\big(v_{2}^{\T}y^{u,\nu}\big)\Big)\,\Bigg]^{\beta}\\
 & =\alpha\kappa_{\eta}(u,v_{1})+\beta\kappa_{\eta}(u,v_{2}).
\end{align*}

\subsection{Linearized connectivity\label{subsec:Linearized-connectivity}}

The effective connectivity $\tilde{W}$ is obtained from linearizing
around the network's time evolution as 
\begin{align*}
(\tau\partial_{t}+1)\,(y_{i}(t)+\delta y_{i}(t)) & =\sum_{j}W_{ij}((y_{j}(t)+\delta y_{j}(t))+\alpha(y_{j}(t)+\delta y_{j}(t))^{2})+u_{i}x(t)\\
\Rightarrow(\tau\partial_{t}+1)\delta y_{i}(t) & =\sum_{j}W_{ij}(1+2\alpha y_{j}(t))\,\delta y_{j}(t)+\mathcal{O}(\delta y^{2})
\end{align*}
and approximating the non-linear system by an equivalent linear one
with connectivity $\tilde{W}_{ij}=W_{ij}(1+2\alpha y_{j}(t))$. The
evolution of the system becomes unstable when the real part of an
eigenvalue $\tilde{\lambda}_{\alpha}$ of the effective matrix $\tilde{W}$
exceeds $1$. The time evolution of $\max_{\alpha}(\text{Re}(\tilde{\lambda}_{\alpha}))$
displayed in \Figref{Responses-and-soft-margin-nonlinear}(a) for
the full system and the $\mathcal{O}(\alpha)$ approximation assures
the stability of the solution and the quality of the approximation.

\subsection{Constrained optimization with Lagrange multipliers\label{subsec:Constrained-optimization}}

We need to optimize \eqref{def_opt_with_Lagrange} 
\[
\mathcal{L}(u,v):=\softM(u,v)+\lambda_{u}(\|u\|^{2}-1)+\lambda_{v}(\|v\|^{2}-1),
\]
where $\softM$ takes the form of \eqref{soft_margin_approx_Gaussian}.
Although the mathematical structure of \eqref{soft_margin_approx_Gaussian}
is simple, the optimization of the expression may present a few pitfalls.
In this section, we describe in detail how to find the projection
vectors given the first four moments of the stimuli.

The linear system can be understood as a special case of the non-linear
system, where some contributions to the soft margin and its gradients
vanish. Therefore, we will distinguish the types of reservoir kernels
only where they are relevant.

\subsection{Prerequisites\label{subsec:Prerequisites}}

The numerical results of the optimization slightly depend on the value
of the control parameter $\eta$ of the soft margin that has to be
fixed. In our examples, with $\eta=10$ the soft margin showed already
very similar extrema as the margin. Smaller values correspond to softer
margins. In practice, a good choice of $\eta$ can be obtained by
comparing for different $\eta$ the optimized readout vector and accuracies
for responses of some reservoir to an arbitrary stimulation. This
procedure is fast and reliable since finding the readout vector for
some $\eta$ is only a quadratic problem. Furthermore, an analysis
of the time evolution of the soft margin for random input projections
can be used as in \figref{Optimization-of-input-output}(b) to achieve
a good estimate of a suitable $\tau$. It can be chosen such that
the soft margin for random stimuli just entered a saturating phase,
so that there is not much improvement expected. Extended phases of
saturation, however, are a sign of forgetting of early parts of the
stimuli in the network and should be avoided.

The main procedure then consists of an alternating optimization of
the input and readout projections. Thereby, we denote $\mathcal{L}(u\vert v)$
as the objective function for input optimization, given $v$, and
$\mathcal{L}(v\vert u)$ analogously. The resulting algorithm (slightly
simplified) is given as pseudocode in algorithm \ref{algo:margin-optimization}.

\begin{algorithm}
\caption{Optimization of \eqref{def_opt_with_Lagrange} using Lagrange multipliers.\label{algo:margin-optimization}}

\begin{algorithmic}[1] \State \Call{Compute}{$\sum_{n}G_{ipn}^{(1)}x_{n}^{\nu}$,
$\sum_{n,m}G_{ipqnm}^{(2)}x_{n}^{\nu}x_{m}^{\nu}$} \For{set of
initial $u$} \Repeat \Procedure{Optimize Input}{$\mathcal{L}(u\vert v)$}
\If{$\|\mu\|=0$} \State \Call{Optimize}{$\kappa_{\eta}\gets\sum_{p,q}u_{p}u_{q}(m_{1pq}-\sigma_{0pq})$}
\Comment{eigenvalue problem} \ElsIf{$\kappa_{\eta}$ far from
saturated} \State \Call{Optimize}{$\kappa_{\eta}\gets\sum_{p}u_{p}m_{0p}+\sum_{p,q}u_{p}u_{q}(m_{1pq}-\sigma_{0pq})$}
\Comment{quadratic problem} \Else \State \Call{Optimize}{$\kappa_{\eta}\gets\sum_{p}u_{p}m_{0p}+\sum_{p,q}u_{p}u_{q}(m_{1pq}-\sigma_{0pq})-\sum_{p,q,r}u_{p}u_{q}u_{r}\sigma_{1pqr}-\sum_{p,q,r,s}u_{p}u_{q}u_{r}u_{s}\sigma_{2pqrs}$}
\EndIf \EndProcedure \Procedure{Optimize Readout}{$\mathcal{L}(v\vert u)$}
\If{$\|\mu\|=0$ and $\alpha=0$} \State \Call{Optimize}{$\kappa_{\eta}\gets-\frac{1}{2}\eta v^{\T}(\Sigma_{0}+\Sigma_{1})v$}
\Comment{eigenvalue problem} \Else \State \Call{Optimize}{$\kappa_{\eta}\gets v^{\T}(M_{0}+M_{1})-\frac{1}{2}\eta v^{\T}(\Sigma_{0}+\Sigma_{1})v$}
\Comment{quadratic problem} \EndIf \EndProcedure \Until{$\kappa_{\eta}$
saturated} \EndFor \end{algorithmic} 
\end{algorithm}

\subsection{Optimization of the input projection\label{subsec:Optimization-of-input}}

The determination of the input projection for fixed readout vector
is best conducted, depending on the situation, by one of three methods
for non-linear kernels and one of two methods for linear ones. The
quantities 
\begin{align*}
m_{0p} & =\sum_{i,n}\,G_{ipn}^{(1)}\,v_{i}\langle\zeta_{\nu}x_{n}^{\nu}\rangle\\
m_{1pq} & =\sum_{i,n,m}\,G_{ipqnm}^{(2)}\,v_{i}\langle\zeta_{\nu}x_{n}^{\nu}x_{m}^{\nu}\rangle\\
\sigma_{0pq} & =\sum_{\substack{i,j,\\
n,m
}
}\,\frac{1}{2}\eta\,G_{ipn}^{(1)}G_{jqm}^{(1)}\,v_{i}v_{j}(\langle x_{n}^{\nu}x_{m}^{\nu}\rangle-\langle\zeta_{\nu}x_{n}^{\nu}\rangle\langle\zeta_{\nu}x_{m}^{\nu}\rangle)\\
\sigma_{1pqr} & =\sum_{\substack{i,j,\\
n,m,o
}
}\,\frac{1}{2}\eta\,(G_{ipn}^{(1)}G_{jqrmo}^{(2)}+G_{iqrmo}^{(2)}G_{jpn}^{(1)})\,v_{i}v_{j}(\langle x_{n}^{\nu}x_{m}^{\nu}x_{o}^{\nu}\rangle-\langle\zeta_{\nu}x_{n}^{\nu}\rangle\langle\zeta_{\nu}x_{m}^{\nu}x_{o}^{\nu}\rangle)\\
\sigma_{2pqrs} & =\sum_{\substack{i,j,\\
n,m,o,l
}
}\,\frac{1}{2}\eta\,G_{ipqmn}^{(2)}G_{jrsol}^{(2)}\,v_{i}v_{j}(\langle x_{n}^{\nu}x_{m}^{\nu}x_{o}^{\nu}x_{l}^{\nu}\rangle-\langle\zeta_{\nu}x_{n}^{\nu}x_{m}^{\nu}\rangle\langle\zeta_{\nu}x_{o}^{\nu}x_{l}^{\nu}\rangle),
\end{align*}
where $G^{(2)}$ is the $\mathcal{O}(\alpha)$ correction of the Green's
function $G^{(1)}$ for linear kernels and $\mu=\langle\zeta_{\nu}x_{n}^{\nu}\rangle$,
$\psi=\langle x_{n}^{\nu}x_{m}^{\nu}\rangle-\langle\zeta_{\nu}x_{n}^{\nu}\rangle\langle\zeta_{\nu}x_{m}^{\nu}\rangle$
and $\chi=\langle\zeta_{\nu}x_{n}^{\nu}x_{m}^{\nu}\rangle$ constitute
the cumulants in the notation (\eqref{Gaussian_distribution_input})
used in the main text. This notation is introduced here and in \subsecref{Optimization-of-readout}
for legibility, although a memory-efficient implementation will compute
only products of Green's functions with stimuli $x^{\nu}$ (for example,
$\mathcal{G}_{ipq\nu}^{(2)}=\sum_{n,m}G_{ipqnm}^{(2)}x_{n}^{\nu}x_{m}^{\nu}$),
which then compose the above abbreviations by performing the averages
over $\nu$. With these abbreviations, the dependence of the soft
margin on the input projection becomes independent of the length of
the input and reads 
\[
\kappa_{\eta}=\sum_{p}u_{p}m_{0p}+\sum_{p,q}u_{p}u_{q}m_{1pq}-\sum_{p,q}u_{p}u_{q}\sigma_{0pq}-\sum_{p,q,r}u_{p}u_{q}u_{r}\sigma_{1pqr}-\sum_{p,q,r,s}u_{p}u_{q}u_{r}u_{s}\sigma_{2pqrs}.
\]

\subsubsection*{Preparations for optimization in non-linear systems}

In the non-linear case, it is advisable to take a few precautions
to reduce computation time and enhance performance. The determination
of the optimal input projection $u$ given a fixed readout projection
$v$ should in the first few, but at least one, iterations neglect
terms of $\mathcal{O}(\alpha)$ and higher in the covariance $\Sigma^{u}$.
In these steps, the soft margin is not strictly optimized, but the
result still yields a good initial guess for the full problem. The
advantage of this procedure is that the computation is much faster
and more likely to achieve a solution near the optimum rather than
some local extremum. In the first steps, the direction of the projection
vector $u$ typically changes rapidly and the quadratic part alone
often has a maximum near the optimum of the full soft margin, as the
neglected terms are at least $\mathcal{O}(\alpha)$. In the readout
optimization, the problem is in general quadratic in case of both
linear and non-linear dynamics, so there is no need to make further
simplifications.

Furthermore, the soft margin is not necessarily convex in $u$ in
the non-linear case and sometimes exhibits plateaus over iteration
steps. We therefore recommend to use a small number of initial projection
vectors, optimize them over a few steps as described below, and then
proceed with the best one after these steps.

\subsubsection*{Case A.}

The simplest case arises if $\|\mu\|=0$, since then $m_{0}$, $\sigma_{1}$
and $\sigma_{2}$ vanish, if one neglects the $\mathcal{O}(\alpha)$
contributions in the non-linear case as discussed above. For normalized
input projections, \eqref{soft_margin_approx_Gaussian} is then maximized
by the eigenvector corresponding to the smallest eigenvalue of $\sigma_{0}-m_{1}$.

\subsubsection*{Case B.}

In the general case where $\|\mu\|\neq0$, it is, as mentioned before,
sometimes helpful to ignore the part related to $\sigma_{1}$ and
$\sigma_{2}$ of the soft margin in the non-linear case to obtain
a good guess of the input projection that maximizes \eqref{def_opt_with_Lagrange}.
Since $m_{1}$, $\sigma_{1}$ and $\sigma_{2}$ vanish when $\alpha=0$,
the same procedure applies in the linear case. The objective then
reads 
\[
\mathcal{L}(u\vert v)\rightarrow u^{\T}m_{0}+u^{\T}m_{1}u-u^{\T}\sigma_{0}u+\lambda_{u}(u^{\T}u-1),
\]
so $u$ and $\lambda_{u}$ are found using 
\begin{align}
\partial_{u}\mathcal{L}=0 & \Rightarrow2(\sigma_{0}-m_{1}-\lambda_{u}\mathbb{I})u=m_{0},\label{eq:soft-margin-gradient-u}\\
\partial_{\lambda_{u}}\mathcal{L}=0 & \Rightarrow u^{\T}u-1=0.\label{eq:soft-margin-gradient-lambda-u}
\end{align}
These equations have many solutions, but for a maximum we further
require negative definiteness of $\partial_{u}^{2}\mathcal{L}\vert_{\lambda_{u}}$.
From this condition follows that $\lambda_{u}<\min\{\sigma\,\vert\,\sigma\text{ is eigenvalue of }\sigma_{0}-m_{1}\}$.
Then, $\sigma_{0}-m_{1}-\lambda_{u}\mathbb{I}$ is symmetric and invertible
and, from solving the first condition for $u$ and inserting in the
second, we get 
\begin{equation}
\frac{1}{4}(m_{0}^{\T}(\sigma_{0}-m_{1}-\lambda_{u}\mathbb{I})^{-1}(\sigma_{0}-m_{1}-\lambda_{u}\mathbb{I})^{-1}m_{0})=1.\label{eq:bisection-problem-lambda-u}
\end{equation}
The term on the left hand side is positive, has poles around the eigenvalues
of $\sigma_{0}-m_{1}$ and deviates only slightly from $0$ for $\lambda_{u}\ll\min\{\sigma\,\vert\,\sigma\text{ is eigenvalue of }\sigma_{0}-m_{1}\}$.
A bisection is therefore best suited to determine $\lambda_{u}$ and
thereby $u$ using \eqref{soft-margin-gradient-u}. This also avoids
running into an undesirable solution where $\|u\|\rightarrow0$ and
$\lambda_{u}\rightarrow-\infty$. However, the poles have only a very
small width and the determination of eigenvalues and inverse matrices
is accompanied by numerical uncertainties. Therefore, the upper bound
on $\lambda_{u}$ is found best as the smallest value within a window
of a small width $\varepsilon$ around the smallest eigenvalue, where
the term on the left hand side exceeds one. Although this corresponds
to a fine-tuning of the Lagrange parameter $\lambda_{u}$ with a sensitive
dependence of the left hand term in \eqref{bisection-problem-lambda-u}
on the exact used eigenvalues, the soft margins corresponding to the
obtained solutions remained robust against neglecting near-vanishing,
and therefore numerically uncertain, eigenvalues in the summation.
Components of the input projection in these directions are neutralized
by their eigenvalues in \eqref{def_opt_with_Lagrange}.

\subsubsection*{Case C.}

If the system is non-linear and a good initial guess for the input
projection is available, predefined solvers, such as the fsolve function
implemented in numpy \citep{oliphant2006guide}, typically find good
solutions for the Lagrange conditions, which are in this case 
\begin{align*}
2(\sigma_{0pq}-m_{1pq}-\lambda_{u}\mathbb{I}_{pq})u_{q}+(\sigma_{1pqr}+\sigma_{1qrp}+\sigma_{1rpq})u_{q}u_{r}\\
+(\sigma_{2pqrs}+\sigma_{2qrsp}+\sigma_{2rspq}+\sigma_{2spqr})u_{q}u_{r}u_{s} & =m_{0p},\\
u^{\T}u & =1.
\end{align*}
The first guess should be the solution from the previous iteration
step. Only if the soft margin reduces by the found solution, a new
guess should be computed neglecting $\sigma_{1}$ and $\sigma_{2}$.
For this comparison, it is important to make sure the projection vectors
are properly normalized. Although this is ensured by the Lagrange
condition, the actual lengths of the returned vectors slightly deviate
from one because of the fine-tuning of the Lagrange parameters. If
the soft margin found near that solution still decreases, we decided
to use the new solution anyway as a restart-point. The readout vector
optimization then improves the soft margin again.

\subsection{Optimization of the readout projection\label{subsec:Optimization-of-readout}}

The optimization of the readout projection is structurally the same
as for the input projection, only the objective function is in general
bi-linear in $v$. The abbreviations used here are 
\begin{align*}
M_{0i} & =\sum_{p,n}\,G_{ipn}^{(1)}\,u_{p}\langle\zeta_{\nu}x_{n}^{\nu}\rangle\\
M_{1i} & =\sum_{\substack{p,q,\\
n,m
}
}\,G_{ipqnm}^{(2)}\,u_{p}u_{q}\langle\zeta_{\nu}x_{n}^{\nu}x_{m}^{\nu}\rangle\\
\Sigma_{0ij} & =\sum_{\substack{p,q,\\
n,m
}
}\,G_{ipn}^{(1)}G_{jqm}^{(1)}\,u_{p}u_{q}(\langle x_{n}^{\nu}x_{m}^{\nu}\rangle-\langle\zeta_{\nu}x_{n}^{\nu}\rangle\langle\zeta_{\nu}x_{m}^{\nu}\rangle)\\
\Sigma_{1ij} & =\sum_{\substack{p,q,r,\\
n,m,o
}
}\,(G_{ipn}^{(1)}G_{jqrmo}^{(2)}+G_{iqrmo}^{(2)}G_{jpn}^{(1)})\,u_{p}u_{q}u_{r}(\langle x_{n}^{\nu}x_{m}^{\nu}x_{o}^{\nu}\rangle-\langle\zeta_{\nu}x_{n}^{\nu}\rangle\langle\zeta_{\nu}x_{m}^{\nu}x_{o}^{\nu}\rangle)\\
 & +\sum_{\substack{p,q,r,s,\\
n,m,o,l
}
}\,G_{ipqmn}^{(2)}G_{jrsol}^{(2)}\,u_{p}u_{q}u_{r}u_{s}(\langle x_{n}^{\nu}x_{m}^{\nu}x_{o}^{\nu}x_{l}^{\nu}\rangle-\langle\zeta_{\nu}x_{n}^{\nu}x_{m}^{\nu}\rangle\langle\zeta_{\nu}x_{o}^{\nu}x_{l}^{\nu}\rangle).
\end{align*}
The objective function to maximize is then 
\[
\mathcal{L}(v\vert u)\rightarrow v^{\T}(M_{0}+M_{1})-\frac{1}{2}\eta v^{\T}(\Sigma_{0}+\Sigma_{1})v+\lambda_{v}(v^{\T}v-1).
\]
Only if the system is linear and the mean stimulus difference $\mu$
is vanishing, this becomes an eigenvalue problem and the optimal readout
vector $v$ is the eigenvector corresponding to the smallest eigenvalue
of $\Sigma_{0}$ (compare case A). Otherwise, the Lagrange parameter
follows from a bisection using the conditions 
\begin{align}
\partial_{v}\mathcal{L}=0 & \Rightarrow(\eta(\Sigma_{0}+\Sigma_{1})-2\lambda_{v}\mathbb{I})v=M_{0}+M_{1},\label{eq:soft-margin-gradient-v}\\
\partial_{\lambda_{v}}\mathcal{L}=0 & \Rightarrow v^{\T}v-1=0.\label{eq:soft-margin-gradient-lambda-v}
\end{align}
From negative definiteness, $\lambda_{v}<\frac{1}{2}\min\{\sigma\,\vert\,\sigma\text{ is eigenvalue of }\eta(\Sigma_{0}+\Sigma_{1})\}$
follows as upper bound on $\lambda_{v}$ (compare case B).

%
%
\end{document}